\pdfoutput=1
\documentclass{webofc}
\usepackage[varg]{txfonts}   
\usepackage{bm}
\usepackage{empheq}
\usepackage{multirow}

\begin{document}
\title{The evaluation of the systematic uncertainties for the finite MC samples in the presence of negative weights}
%
%

\subtitle{Presented at The XXIII International Workshop \\``High Energy Physics and Quantum Field Theory''}

\author{\firstname{Petr} \lastname{Mandrik}\inst{1}\fnsep\thanks{\email{Petr.Mandrik@ihep.ru}}
}

\institute{NRC «Kurchatov Institute» – IHEP, Protvino
          }

\abstract{%
  The analysis of results from HEP experiments often involves the estimates of the composition of the binned data samples, based on Monte Carlo simulations of various sources. Due to a finite statistic of MC samples they have statistical fluctuation. This work proposes the method of incorporating the systematic uncertainties due to finite statistics of MC samples with negative weights. The possible approximations are discussed and the comparison of different methods are presented.
}
\maketitle
%

\section{Introduction} \label{intro}

Experimental results in high energy physics are often represented as a binned distribution (histogram) of observed events $\bm{X} = (X_1, X_2, ... )$, where $X_i$ is a number of events in bin $i$.
The usual method to estimate physical parameters such as particle masses or cross sections from this distribution is to 
perform some of Bayesian or frequentist analyses based on likelihood function. Likelihood function $\mathcal{L}(\bm{X}|\bm{m})$ connect the data with a theoretical model and represente how well the observatios are described by the prediction $\bm{m} = (m_1, m_2, ... )$ . This prediction may depend on several different parameters $\bm{\pi} = (\pi_1, \pi_2, ...)$: nuisance parameters and parameters of interests. In addition if some signals or background processes are known from Monte-Carlo simulations then the likelihood function depends on template distributions $\bm{t} = (t^a_1, t^a_2, ... t^b_1, t^b_2, ... t^c_1, t^c_2, ... )$, where $t^k_i$ is a number of events for process $k$ in bin $i$:
\begin{equation} \label{likelihood}
  \mathcal{L}(\bm{X}|\bm{m}) = \mathcal{L}(\bm{X}|\bm{\pi}, \bm{t}) = \prod_i P(X_i|\bm{\pi}, \bm{t}_i) = \prod_i P(X_i|\pi_1, \pi_2, ..., t^a_i, t^b_i, ...)
\end{equation}

This is important task to define an adequate likelihood function and take into account all existed statistical and systematic uncertainties present in the analysis.

Template distributions from Monte-Carlo generators are subject to statistical fluctuations due to finite number of events in samples.
The influence of these fluctuations can be expected to be significant in regions of low amounts of Monte-Carlo events.
For incorporating such uncertainties into likelihood function Barlow and Beeston proposed a method \cite{Barlow:1993dm} wherein one for every bin $i$ and every process $k$ introduces a new parameter $T_i^k$ corresponding to unknown expected number of events in infinite statistics limit:
\begin{equation} \label{barlow}
  \prod_i P(X_i|\bm{\pi}, \bm{t}_i) \rightarrow 
  \prod_i \Big[ P(X_i|\bm{\pi}, \bm{T}_i) \cdot \prod_k P(t^k_i|T^k_i) \Big]
\end{equation}
where for constrain $P(t^k_i|T^k_i)$ Barlow and Beeston assumed a Poisson distribution.

On the other hand several of the modern Monte-Carlo generators \cite{Buckley:2011ms} produce a weighted events with both negative and positive weights. In this case the transformation (\ref{barlow}) is not applicable.
In this paper we provide a method of incorporating uncertainties due to the finite statistics of Monte-Carlo samples in the presence of negative weights.

\section{Likelihood functions for Monte-Carlo samples with negative weights} \label{sec-1}

In simplified form the algorithm of event production in most important example of Monte-Carlo generator with negative weights MadGraph5\_aMC@NLO can be described as follow \cite{Frixione:2002ik}. A cross section of some process $\sigma_{NLO}$ is calculated by computing the integrals of two functions $F_H(x)$ and $F_S(x)$:
\begin{equation}
  \sigma_{NLO} = \int F_H(x) dx + \int F_S(x) dx
\end{equation}

By definition, the functions $F_H(x)$ and $F_S(x)$ are finite and $F_H(x)+F_S(x)>0$, but for some values of $x$ the function $F_S(x)$ is negative.
Using the absolute values of the integrands $|F_H(x)|$ and $|F_S(x)|$ two set of events are produced - $\{x\}_H$ and $\{x\}_S$ respectly with weights $w_i^{H,S}$ equal to $+1$ for positive values of functions and weight $-1$ if function is negative, so:
\begin{equation} \label{mad_xces}
  \sigma_{NLO} = \frac{\int|F_H(x)| dx}{N_H} \cdot \sum_i^{N_H} w_i^H + \frac{\int|F_S(x)| dx}{N_S} \cdot \sum_i^{N_S} w_i^S
\end{equation}
where $N_H$, $N_S$ are thw number of events in corresponding sets.

In this way in the infinite statistics limit the prediction of any observable in any intervals $[x_i, x_i + \Delta x]$ of histograms from MadGraph5\_aMC@NLO can be only positive, but in the case of finite statistics the prediction could get negative values in some bins.
On the other hand the events with negative and positive weights should be treated in the same way during analyses and pass the same cuts to keep the correct cross section value (\ref{mad_xces}). Further we assume that the last condition is satisfied, so for the finite number of generated Monte-Carlo events the probability of obtaining a given one in the case of only positive or negative weights is described by multinomial distribution, which is usually approximated by multiplication of independent Poisson distributions (see for example, \cite{Walck:1996cca}).

Let us consider a simple case of single bin and only one generated process with total number of event $t$.
If $t^{+}$ is a sum of all positively weighted events from Monte-Carlo samples and $t^{-}$ is a sum of all negative,
then $t$ is a difference of two Poissonian quantities $t^{+}$ and $t^{-}$ and described by Skelleman distribution:
\begin{multline} \label{skelleman}
  P(t) = \mathcal{S}(t | T^+, T^-) = \sum_{s = max(0,t)}^{\infty} \mathcal{P}(s|T^+) \cdot \mathcal{P}(s - t|T^-)
  = e^{-(T^+ + T^-)} \Big( \frac{T^+}{T^-} \Big)^{t/2} \mathcal{I}_t(2\sqrt{T^+ T^-})
\end{multline}
where $\mathcal{I}_t$ is a modified Bessel functions of the first kind, $\mathcal{P}$ is Poisson distribution, $T^+$ and $T^-$ are the parameters of Skellenam distribution, corresponding to unknown ``true'' prediction of negative and positive events from MC generator.

Using the equations (\ref{skelleman}) in (\ref{barlow}) we obtain the following transformation rule for taking into account uncertainties due to the finite statistics of Monte-Carlo samples in the presence of negative weights in likelihood function:
\begin{equation}  \label{barlow_skelleman}
  \prod_i P(X_i|\bm{\pi}, \bm{t}_i) \rightarrow
  \prod_i \Big[ P(X_i|\bm{\pi}, \bm{T}_i) \cdot \prod_k \mathcal{S}(t^k_i|T^{k+}_i, T^{k-}_i) \Big]
\end{equation}
where the new parameters are related by the equation $T^k_i = T^{k+}_i - T^{k-}_i$.

The constrain on parameters $T^k_i$, $T^{k+}_i$, $T^{k-}_i$ in formula (\ref{barlow_skelleman}) may be improved by an independent treatment of 
values $t^{+}$ and $t^{-}$ in analyses. In this case we get:
\begin{equation}\label{skelleman_forvard}
  P(t) = \mathcal{P}(t^{+}|T^+) \cdot \mathcal{P}(t^{-}|T^-)
\end{equation}
and from (\ref{barlow}) with (\ref{skelleman_forvard}):
\begin{equation} \label{barlow_skelleman_forvard}
  \prod_i P(X_i|\bm{\pi}, \bm{t}_i) \rightarrow 
    \prod_i \Big[ P(X_i|\bm{\pi}, \bm{T}_i) 
    \cdot \prod_k \mathcal{P}(t^{k+}_i|T^{k+}_i) \cdot \mathcal{P}(t^{k-}_i|T^{k-}_i) \Big]
\end{equation}

The number of extra parameters in the transformation rule (\ref{barlow_skelleman_forvard}) is equal to $2\times$ \textit{number of processes} $\times$ \textit{number of bins}.
We can decrease the number of parameters by using the method of maximum likelihood function
Indeed, if $\mathcal{L}$ is a likelihood function with transformation (\ref{skelleman_forvard}) then for bin $i$ one gets:
\begin{multline}
  -\ln{ \mathcal{L}_i } = -\ln{P(X_i|\bm{\pi}, \bm{T}_i)}
  - \sum_k \Big( -T^{k+}_i + t^{k+}_i \cdot \ln{T^{k+}_i} - \ln{(t^{k+}_i!)} \Big) - \\
  - \sum_k \Big( -T^{k-}_i + t^{k-}_i \cdot \ln{T^{k-}_i} - \ln{(t^{k-}_i!)} \Big)
\end{multline}

The requirement of an extremum gives the following system of equations:
\begin{empheq}[left = \empheqlbrace]{align} \label{system_of_eq}
  \begin{split}
  -\frac{ \partial \ln \mathcal{L} }{ \partial T^{k+}_i } 
    & = 1 -\frac{ \partial \ln P(X_i|\bm{\pi}, \bm{T}_i) }{ \partial T^{k+}_i } 
    - \frac{t^{k+}_i}{T^{k+}_i} = 0\\ 
   -\frac{ \partial \ln \mathcal{L} }{ \partial T^{k-}_i } 
    & = 1 -\frac{ \partial \ln P(X_i|\bm{\pi}, \bm{T}_i) }{ \partial T^{k-}_i } 
    - \frac{t^{k-}_i}{T^{k-}_i} = 0
\end{split}
\end{empheq}
This system (\ref{system_of_eq}) in some cases may be solved analytically for the parameters $T^{k-}_i$, $T^{k+}_i$, 
or numerically with some fixed values of the remaining parameters.

The another way to decrease the number of parameters related to finite statistics of Monte-Carlo is known as Barlow-Beeston ``light'' transformation \cite{Conway:2011in}. As the statistical uncertainties for each source in each bin are independent they may be combined and be represented approximately by single effective parameter per bin $M_i$:
\begin{equation} \label{conway}
  \prod_i P(X_i|m_i) \rightarrow 
  \prod_i P(X_i|M_i) \cdot P(m_i|M_i)
\end{equation}
Usually, in this approximation for $P(m_i|M_i)$ usually a Gaussian constrain  $\mathcal{G}(m_i|M_i, \sigma_i)$ is used, where
the value of $\sigma_i$ are calculated by propagation of the Monte-Carlo statistical uncertainties in bin $i$ with fixed values of the remaining parameters.

For histograms with negative weights the transformation (\ref{conway}) has the form:
\begin{equation} \label{conway_nega}
  \prod_i P(X_i|m_i^+ - m_i^-) \rightarrow 
  \prod_i P(X_i|M_i^+ - M_i^-) \cdot P(m_i^+|M_i^+) \cdot P(m_i^-|M_i^-)
\end{equation}
The number of extra parameters is equal to $2 \times$ \textit{number of bins in histogram}.

For the likelihood function with transformation (\ref{conway_nega}) a system of equations similar to (\ref{system_of_eq}) can be obtained. Using the Gaussian constrain one gets:
\begin{equation}
  \mathcal{L}_i = \mathcal{P}(X_i|M_i^+ - M_i^-) \cdot \mathcal{G}(m_i^+|M_i^+, \sigma_i^+) \cdot \mathcal{G}(m_i^-|M_i^-, \sigma_i^-)
\end{equation}
\begin{multline}
-\ln{ \mathcal{L}_i } = -\Big[-(M_i^+ - M_i^-) + X_i \cdot \ln{(M_i^+ - M_i^-)} - \ln{X_i!}\Big] - \\
                        -\Big[ \frac{(M_i^+ - m_i^+)^2}{2(\sigma_i^+)^2} - \ln{\sigma_i^+ \sqrt{2 \pi }} \Big]
                        -\Big[ \frac{(M_i^- - m_i^-)^2}{2(\sigma_i^-)^2} - \ln{\sigma_i^- \sqrt{2 \pi }} \Big]
\end{multline}
\begin{empheq}[left = \empheqlbrace]{align} \label{system_of_eq_conw}
 \begin{split}
  -\frac{ \partial \ln \mathcal{L} }{ \partial M_i^+ } 
  & = 1 - \frac{X_i}{M_i^+ - M_i^-} - \frac{M_i^+ - m_i^+}{(\sigma_i^+)^2} = 0 \\
  -\frac{ \partial \ln \mathcal{L} }{ \partial M_i^- } 
  & = 1 + \frac{X_i}{M_i^+ - M_i^-} - \frac{M_i^- - m_i^-}{(\sigma_i^-)^2} = 0
\end{split}
\end{empheq}

\section{The performance of methods}

In this section few results of study the proposed transformations for taking into account uncertainties due to the finite statistics of Monte-Carlo samples are given.
The source code was implemented with statistical package SHTA \cite{shta2017}.

From the simple single-bin single-channel model:
\begin{equation} \label{test_model_0}
  \mathcal{L}_{0} = \mathcal{P}(X|\pi \cdot (C + T^+ - T^-)) \cdot \mathcal{P}(t^+|T^+) \cdot \mathcal{P}(t^-|T^+)
\end{equation}
a set of events $(X, t^+, t^-)$ may be generated for the fixed values of parameters $\pi$, $T^+$, $T^-$.
Here the constant $C$ is introduced in order to avoid a long tail in posterior distribution of $\pi$ from $T^+ - T^- \sim 0$.

To estimate the parameter of interest $\pi$ we use three different likelihood functions.
First of all a naive approach without incorporating uncertainties due to the finite statistics of Monte-Carlo samples:
\begin{equation} \label{check_model_0_n}
  \mathcal{L}_{n} = \mathcal{P}(X|\pi \cdot (C + t^+ - t^-))
\end{equation}
a likelihood function with transformation (\ref{barlow_skelleman_forvard}):
\begin{equation} \label{check_model_0_p}
  \mathcal{L}_{p} = \mathcal{P}(X|\pi \cdot (C + T^+ - T^-)) \cdot \mathcal{P}(t^+|T^+) \cdot \mathcal{P}(t^-|T^-) \cdot \mathcal{H}(T^+-T^-)
\end{equation}
and similar one but with Gaussian approximation for multiplication of two Poissons:
\begin{equation} \label{check_model_0_g}
  \mathcal{L}_{g} = \mathcal{P}(X|\pi \cdot (C + T)) \cdot \mathcal{G}(t^+-t^-, T, \sqrt{t^+ + t^-}) \cdot \mathcal{H}(T)
\end{equation}
where $\mathcal{H}$ is a Heaviside function.

The generated set of toy data from (\ref{test_model_0}) is used to perform a Bayesian inference (see for example \cite{DAgostini:1999gfj}) with non-informative flat priors for all parameters.
The posterior probability density functions for parameters were obtained from likelihood functions (\ref{check_model_0_n}), (\ref{check_model_0_g}), (\ref{check_model_0_p}) and the confidence intervals were found.
The results for two different set of initial values of $\pi$, $T^+$, $T^-$ are presented in the table \ref{table:results_0}.

\begin{table}
\centering
\caption{ Percent of toy data for which the confidence interval with confidence level CL from posterior distribution includes the true value of parameter $\pi$. The uncertainties are evaluated by averaging the results from different sets of toy data.}
\label{table:results_0}       
\begin{tabular}{|c|c|c|c|c|}
\hline
      $\mathcal{L}_{0}$ parameters & CL    & $\mathcal{L}_{n}$    & $\mathcal{L}_{g}$    & $\mathcal{L}_{p}$  \\
      \hline
      \multirow{2}{*}{$\pi = 3, T^+ = 12, T^- = 4$} &  $1\sigma$ & $34.71 \pm 0.79$ & $67.79 \pm 0.75$ & $67.98 \pm 0.72$ \\ 
                             &  $2\sigma$ & $62.82 \pm 0.70$ & $95.25 \pm 0.27$ & $96.14 \pm 0.22$ \\ 
      \hline
      \multirow{2}{*}{$\pi = 3, T^+ = 9, T^- = 7$} &  $1\sigma$ & $26.41 \pm 0.57$ & $77.52 \pm 0.63$ & $78.51 \pm 0.63$ \\ 
                             &  $2\sigma$ & $49.73 \pm 0.71$ & $98.18 \pm 0.15$ & $98.21 \pm 0.15$ \\ 
      \hline
\end{tabular}
\vspace*{0.2cm}  
\end{table}

From the table \ref{table:results_0} we can see that the difference between solution with multiplication of two Poisson and its Gaussian approximation do not exceed one percent.
On the other hand without taking into account the uncertainties due to the finite statistics of Monte-Carlo samples in likelihood function (\ref{check_model_0_n})
the accuracy of measurements falls significantly.
For example, only in two out of three experiments the correct interval will be obtained for $CL = 2\sigma$ and first set of parameters.


\section{Conclusion}
In this work a method of incorporating the systematic uncertainties due to finite statistics of MC samples with negative weights is presented.
The influence of this statistical uncertainty can be expected to be high in regions of low amounts of Monte Carlo events and they must be included into the fit.
The proposed transformation (\ref{barlow_skelleman_forvard}) and its simplified version (\ref{conway_nega}) can be used to construct the correct likelihood function.
While using the Gaussian approximation of multiplication of two Poisson distribution in (\ref{barlow_skelleman_forvard}) or (\ref{conway_nega}) leads to the known expressions used in different 
statistical packages \cite{thetaarticle}\cite{Cranmer:1456844} in different forms, the choice of specific form of likelihood function depends on the analysis and in some cases the more accurate proposed methods can improve the results.

\section*{Acknowledgments}
I wish to thank L. Dudko S. Slabospitskii and G. Vorotnikov for useful discussions. 

\bibliography{nega_text_eng}
%
%
%
%

\end{document}